\documentclass[aps,twocolumn]{revtex4}

\usepackage{graphicx}
\usepackage{dcolumn}
\usepackage{bm}
\usepackage{amsmath}
\begin{document}

\newcommand{\bk}{{\bf k}}
\newcommand{\bp}{{\bf p}}
\newcommand{\bv}{{\bf v}}
\newcommand{\bq}{{\bf q}}
\newcommand{\tbq}{\tilde{\bf q}}
\newcommand{\tq}{\tilde{q}}
\newcommand{\bQ}{{\bf Q}}
\newcommand{\br}{{\bf r}}
\newcommand{\bR}{{\bf R}}
\newcommand{\bB}{{\bf B}}
\newcommand{\bE}{{\bf E}}
\newcommand{\bA}{{\bf A}}
\newcommand{\bK}{{\bf K}}
\newcommand{\vd}{{v_\Delta}}
\newcommand{\tr}{{\rm Tr}}

\title{Majorana Fermions: the race continues}

\author{M. Franz}
\affiliation{Department of Physics and Astronomy,
University of British Columbia, Vancouver, BC, Canada V6T 1Z1}
\date{\today}

\begin{abstract}
\end{abstract}
\maketitle

When in 1937 Ettore Majorana discovered a purely real-valued solution 
\cite{majorana} to the celebrated Dirac equation, he could not have foreseen the
whirlwind of activity that would follow -- some 70
years later  -- and  not in particle physics, that was his domain but in
nanoscience and condensed matter physics.  Majorana fermions, as the
particles described by these solutions became known, are curious
objects. The recent storm of activity in condensed matter physics has
focused on the `Majorana zero modes' i.e.\ emergent Majorana particles occurring at
exactly zero energy that have a remarkable property  of being their
own antiparticles \cite{wilczek1,franz1}. Mathematically, this
property is expressed as an equality between the particle's creation
and annihilation operators,   $\gamma^\dagger=\gamma$. As explained more fully below any ordinary fermion can be though of as composed of two Majorana fermions. An interesting situation arises when a {\em single} Majorana particle can be spatially separated from its partner and independently probed.  Observation of such an  `unpaired' Majorana particle in a solid-state system would clearly fulfill a  
longstanding intellectual challenge. In addition, Majorana zero
modes are believed to exhibit the so called non-Abelian exchange statistics \cite{stern1,nayak1} which endows them with a technological potential as building blocks of future quantum memory immune against many sources of
decoherence which plague other such proposed devices.

Recent advances in our understanding of solids with strong spin-orbit coupling,
combined with the progress in nanofabrication, put the physical
realization of the Majorana particles within reach. In fact signatures
consistent with their existence in quantum wires coupled to conventional
superconductors in a setup schematically depicted in Fig.\ 1 have been
 reported by several groups
\cite{mourik1,larsson1,heiblum1,harlingen1,marcus1,rokhinson1}. The theory behind these
devices is very well understood -- it is rooted in the standard band
theory of solids and the Bardeen-Cooper-Schrieffer theory of
superconductivity -- and there is no doubt that Majorana zero modes
should appear {\em under the right conditions}. The key question that
remains to be answered is this: Have the right conditions been
achieved in the existing devices? It is this question that we address
in this Commentary after presenting a brief overview of the field.
\begin{figure}
\includegraphics[width =8cm]{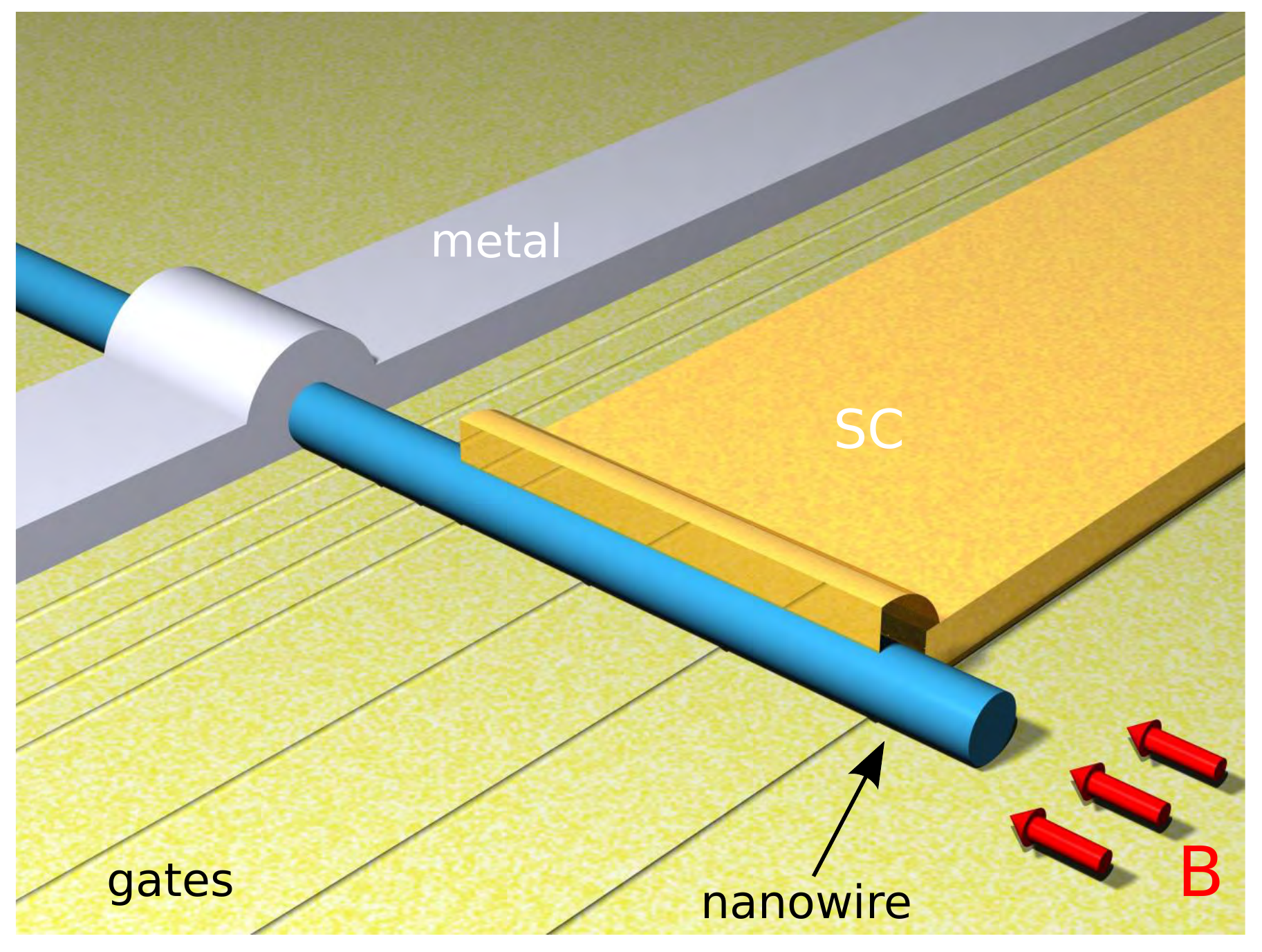}
\caption{A typical experimetal setup for the Majorana zero mode detection in a nanowire. The nanowire is placed on a substrate equipped with gates and contacted from above by a superconducting (SC) and normal metal electrodes. 
}
\end{figure}

\section{Why wires?}
Historically, there has been a number of proposals to engineer and detect
Majorana zero modes in two-dimensional solid state systems, including the
fractional quantum Hall liquids \cite{read1},  interacting quantum
spin systems
\cite{kitaev2}, spin-polarized $p$-wave superconductors \cite{read2},
and more recently interfaces between topological insulators \cite{fu2}
or semiconductors \cite{sau1,alicea1} and ordinary superconductors. Despite significant progress, especially in the quantum Hall liquids, none of these proposals has seen a decisive experimental confirmation. Instead, the focus over the past two years has shifted to one-dimensional
structures -- quantum wires -- which are thought to posses several
distinct advantages when it comes to fabrication and subsequent
detection of the Majorana zero modes. As explained more fully below,
in quantum wires Majoranas occur either at the wire end or at a
domain wall between topological and non-topological regions of the wire. This
facilitates a relatively easy detection compared to 2D systems where
Majoranas live in the cores of magnetic vortices  or other
topological defects, which can be located essentially anywhere in the
sample and thus difficult to find. The second key 
advantage is the expected paucity, relative to
the 2D systems, of various low-energy excitations 
that could interfere with the detection of Majorana zero
modes. This is due to the 1D confinement of electronic and other
degrees of freedom in the wire geometry. Finally, there have been
significant recent advances in fabrication
and manipulation of clean quantum wires which allow for an unprecedented
level of control and analysis in a wide range of settings.    

The prototype on which all the current 1D devices, both proposed and
realized, are based is the Kitaev chain \cite{kitaev1} illustrated in
Fig.\ 2. The model, elegant in its simplicity, describes spinless
electrons hopping between the sites of a 1D tight-binding chain and subject to
superconducting pairing with $p$-wave symmetry (i.e.\ on the bonds
connecting the neighboring sites). Kitaev observed that depending on
the model parameters, the hopping amplitude $t$ and the SC pairing
amplitude $\Delta$, the system can be in two distinct phases. If we
think of the fermions on each lattice site as composed of two Majorana fermions,
$c_j=\gamma_{j1}+i\gamma_{j2}$, then the trivial phase can be depicted
as in the top panel of Fig.\ 2, and has Majoranas bound into ordinary fermions on each
site. In the other phase, Majoranas on the {\em neighboring} sites
bind to form a regular fermion, leaving an unpaired Majorana at each
end of the chain as illustrated in the bottom panel of Fig.\ 2.  This is the topological SC
phase that underlies all the recent proposals to engineer Majorana
fermions in 1D devices.

\section{Physical realizations of the Kitaev chain}

There exist two basic realizations of the Kitaev chain. One is based
on quantum wires made of a semiconductor with strong spin-orbit coupling such
as InSb or InAs and the other employs wires made of a 3D topological insulator (TI)
such as Bi$_2$Se$_3$. In both cases superconductivity in the wire must
be induced via the proximity effect in a setup schematically depicted in
Fig.\ 1.  Magnetic field is used to produce effectively spinless
electrons in the wires as required by the Kitaev paradigm. 
\begin{figure}
\includegraphics[width = 8cm]{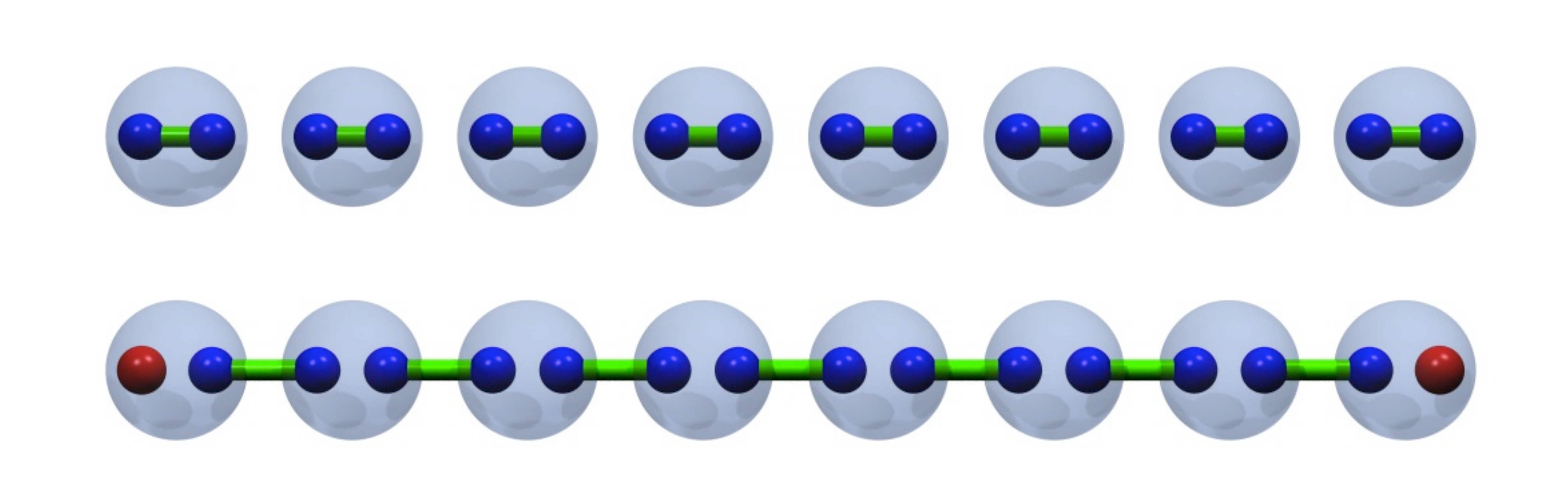}
\caption{Two phases of the Kitaev chain. Trivial phase (top) has Majorana fermions (blue spheres) bound in pairs located on the same site of the physical lattice, represented by translucent spheres. In the topological phase (bottom) Majorana femions are bound in pairs located on the neighboring sites leading to two unpaired Majoranas at both ends, represented by the red spheres.
}
\end{figure}

We discuss first the semiconductor wire implementation
\cite{lutchyn1,oreg1} that has
attracted by far the most attention over the past two years. The well
known excitation spectrum of such a wire is illustrated in Fig.\ 3a:  the Rashba
spin-orbit coupling separates the parabolic bands for two spin
projections and the Zeeman field opens up a gap $V_Z=gB$ near $k=0$ leading to an
effectively spinless 1D metal when the chemical potential $\mu$ lies in the
Zeeman gap. This is the key condition that must be met in order to
realize a 1D topological superconductor; more generally one requires
an {\em odd} number of Fermi points in the right half of the Brillouin
zone \cite{kitaev1}. Superconducting order in such a wire gives rise to the
topological phase when the following condition on the superconducting
gap magnitude $\Delta$ is satisfied: $V_Z>\sqrt{\Delta^2+\mu^2}$. In
this regime the semiconductor wire realizes the Kitaev chain paradigm and will have Majorana zero modes localized at its ends. 

The condition on $V_Z$, $\Delta$ and $\mu$ listed above imposes some
considerable constraints on the physical realization of the
topological phase \cite{alicea2}. For typical values of
the magnetic $g$-factor (15 and 50 for InAs and InSb wires, respectively)
and for the magnetic fields of few Tesla one obtains $V_Z\simeq 1-10$K. 
Tuning the
chemical potential $\mu$ with this accuracy and ensuring that it is
also sufficiently homogeneous so that 
the condition is satisfied everywhere along the length of the wire  represents a
significant experimental challenge. Also, the smallness of $V_Z$ restricts the experimental window for Majorana fermion observation and manipulation to low temperatures $T\ll V_Z$. Yet, several groups have reported signatures in wires consistent with the existence of Majorana zero modes \cite{mourik1,larsson1,heiblum1,rokhinson1,harlingen1,marcus1}. If true, this is a remarkable achievement, although as we discuss below there exist alternative interpretations of these experiments that do not involve Majorana zero modes.

The above mentioned constraint is relaxed in quantum wires
made of a 3D topological insulator \cite{cook1,cook2}. The underlying
physics here is quite different and relies on the topologically
protected surface states that are the halmark of these remarkable
materials \cite{moore_rev,kane_rev}. It is easy to show by an explicit
calculation \cite{wormhole} that the spectrum of such surface states
in a wire whose cross-section is threaded by magnetic flux
$(n+{1\over2})\Phi_0$, with $n$ an integer and $\Phi_0=hc/e$ the
magnetic flux quantum, has the form illustrated in Fig.\ 3b. It
consists of a pair of non-degenerate linearly dispersing gapless modes and a set of doubly degenerate gapped modes. The important property of this spectrum is that the number of Fermi points in the right half of the Brillouin
zone is odd for {\em any} value of the chemical potential as long as
it lies inside the
bulk bandgap, which is $\sim 300$meV in Bi$_2$Se$_3$ family of
materials. Thus, such a wire conforms to Kitaev's paradigm and will
exhibit Majorana zero modes when superconducting order is induced in
it by the proximity effect \cite{cook1}. In addition, unlike in the
semiconductor wires, superconducting order in this setup is expected
to be robust against the effects of non-magnetic disorder
\cite{cook2}. 

As of this writing the existence of coherent surface states in TI
wires has been established \cite{peng1} and the superconducting
proximity effect has been demonstrated
\cite{zhangnanoribbon}. However, signatures of Majorana zero modes
have not yet been reported. The key difficulty appears to lie in the fact that
as in most bulk TIs the chemical potential in the wires is pinned in the conduction
band thus obscuring the universal physics of the surface modes.
\begin{figure}
\includegraphics[width = 8cm]{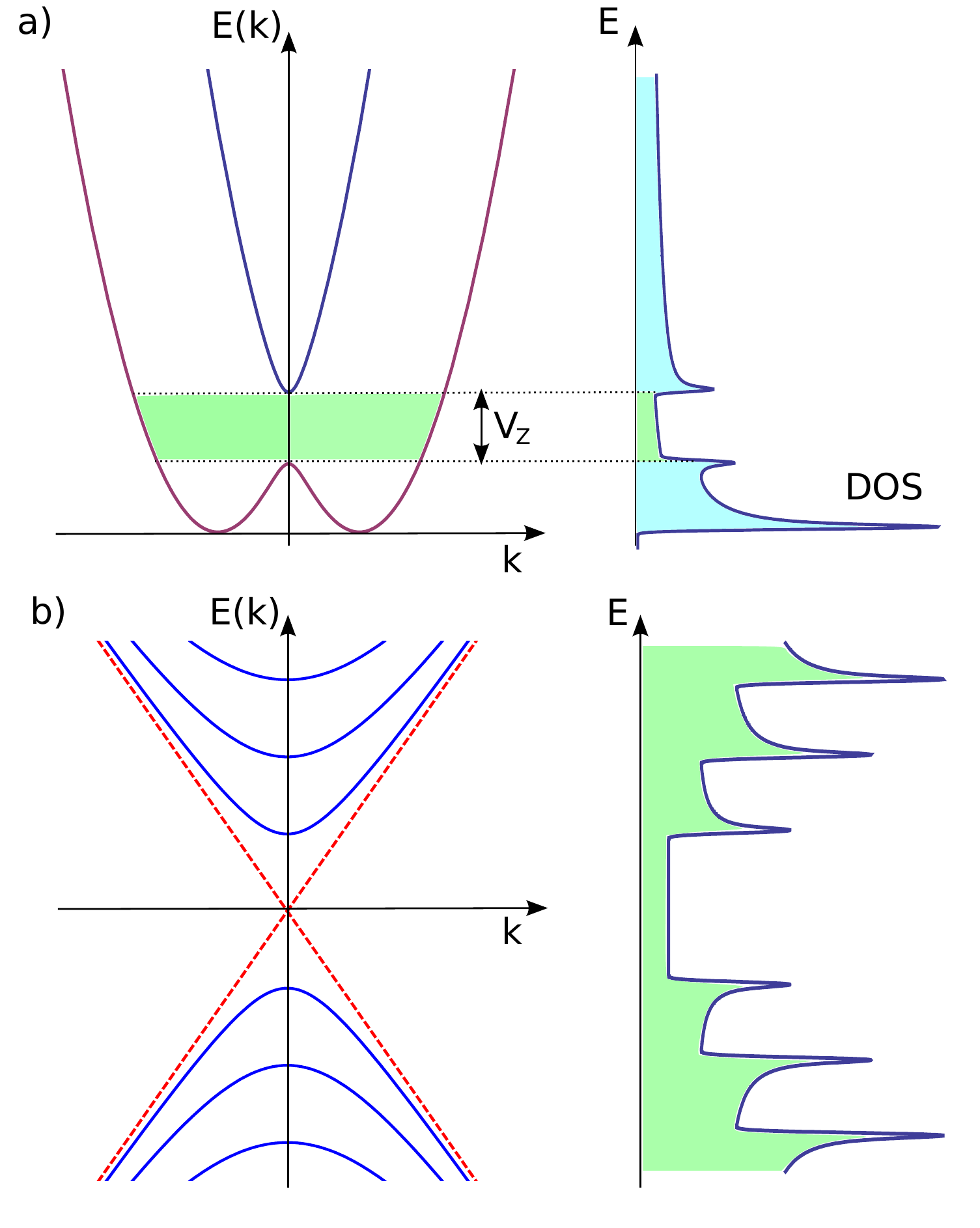}
\caption{Electron excitation spectra in semiconductor and TI nanowires. 
a) The characteristic spectrum of a semiconductor wire with Rashba and Zeeman coupling. In order to achieve the topological phase the chemical potential $\mu$ must lie in the green-shaded region of width $V_Z$. b) The spectrum of a TI wire in parallel magnetic field. The red dashed lines correspond to non-degenerate bands while the blue solid lines are doubly degenerate. 
The right panels show the corresponding density of states (DOS)
}
\end{figure}

\section{Majorana or not?}
The easiest way to experimentally observe the Majorana zero modes in
wires is through the tunnelling spectroscopy. The tunnelling conductance
$dI/dV$ into the wire's end should show a superconducting gap with a peak at the zero bias
when the Majorana mode is present and no peak when it is absent. In
addition, under ideal conditions (i.e.\ low temperature and weak
disorder), the zero-bias peak conductance should be quantized \cite{beenakker1} at
$2e^2/h$  and the SC gap should exhibit a closing at the phase transition
between  the topological and non-topological phase. The existing
experiments \cite{mourik1,larsson1,heiblum1,harlingen1,marcus1} performed in a
setup similar to Fig.\ 1 generally observe a `soft SC gap' (i.e.\
reduced but non-zero density of states at low bias) with a
non-quantized zero-bias peak emerging  and then disappearing as the
magnetic field is increased. The zero-bias peak is interpreted as an
evidence for the Majorana zero mode.  There is typically no sign of SC 
gap closing associated with the appearance of the zero-bias peak. This
last feature at least can be understood from theoretical models
\cite{sau77} which tend to show unambiguous gap closing for tunnelling
into the bulk of the wire but, under a wide range of conditions, no
apparent gap  closing for tunnelling into the wire end. A control
experiment showing $dI/dV$ for both the middle and the end of the wire
would thus be desirable to resolve this particular issue.

The big question of course remains whether or not the observed non-quantized
zero-bias peaks \cite{mourik1,larsson1,heiblum1,harlingen1,marcus1} reflect
the Majorana zero mode. Recent theoretical work \cite{lee1} showed
that in the presence of disorder such non-quantized zero-bias peaks generically
appear even when the wire is in the non-topological phase. 
These  do not correspond to the Majorana modes but
instead to ordinary Andreev bound states that exist close to zero
energy. To make things more complicated it turns out \cite{lee1}  that
such Andreev states can have similar dependence on the applied
magnetic field as the Majorana modes so the experimentally observed
appearance and disappearance of the peaks as a function of $B$ does
not really constitute an unambiguous evidence for the Majorana modes, as was
originally thought. 

Another class of experiments tests the $4\pi$-periodic Josephson
effect that is predicted to occurr between two SC wires in the
topological phase \cite{kitaev1}. Josephson effect between two ordinary
superconductors involves Cooper pair tunnelling and the current shows
the characteristic $I(\varphi)=I_c \sin{\varphi}$ behavior which is
$2\pi$-periodic in the relative phase variable $\varphi$. For two
topological  superconductors the existence of
the Majorana zero modes enables single-electron tunnelling, which
introduces a $\sin{(\varphi/2)}$ component to the tunnelling current.
Ref.\ \cite{rokhinson1} probed this effect by measuring the Shapiro
steps which occur when a junction is exposed to an AC electromagnetic
field. The $4\pi$-periodic Josephson component is manifested by the doubling of
the Shapiro step height which was indeed observed when the wires were
subjected to a static magnetic field. 
The $4\pi$-periodic Josephson effect was thought to
represent a more reliable signature of Majorana fermions in this
context than the tunnelling spectroscopy. Recently, however, it was
shown theoretically \cite{halperin1} that the $4\pi$-periodic Josephson effect can
actually arise under certain conditions even for Josephson junctions
formed of ordinary superconductors with no Majorana zero modes.  Thus,
the experimental result \cite{rokhinson1} weighs in favor of Majorana
interpretation but does not constitute a definitive proof.

\section{The road ahead}
Theoretical models provide rather unambiguous predictions for the
existence of Majorana zero modes in nanowires made of semiconductors
with strong spin orbit coupling and of topological insulators.  The
experimental data currently available show features broadly consistent
with the expectations for Majorana fermions in semiconductor wires but
a number  of puzzles and challenges remain. Among
these perhaps the most vexing is that it has been almost too easy
to observe these elusive modes. Given what we know about the purity of
the wires, the interfaces, and the effect of gating it should have
been more difficult to tune
the chemical potential to lie inside the Zeeman gap that is just few
Kelvins wide. This is
illustrated in the density of states plot in Fig.\ 3a: it shows
that if the electron density  in the wire is set at random (as  would
be the case for a wire in direct contact with a superconductor) the
chances for the chemical potential accidentally landing in the gap are
rather small. 

Nevertheless the data by several different groups
\cite{mourik1,larsson1,heiblum1,harlingen1,marcus1}  using variants of
the basic setup show  the (non-quantized) zero-bias peak which appears
in the range of magnetic fields consistent with the Majorana
prediction. What does this mean? It is possible that for some unknown
reason the Majorana zero modes are more stable and occur under wider
range of conditions than theoretically expected. It is possible that
we have been lucky in terms of achieving the necessary conditions  in the existing devices. It is also possible that the
non-quantized zero-bias peaks reflect ordinary Andreev states instead
of Majorana zero modes \cite{lee1}. Then there is the $4\pi$-periodic
Josephson effect \cite{rokhinson1} which now appears to be a necessary
but not a unique consequence of the Majorana zero modes
\cite{halperin1} adding into the evidence. 

Where do we go from here? It is now thought that an observation of a stable
quantized $2e^2/h$ zero bias peak would constitute a smoking gun proof
of the Majorana zero mode;  it is generally held true that mimicking
this type of
exact quantization would be difficult. Experimentally, such an
observation using present day devices would require an unrealistically low
temperatures but can be perhaps achieved in future devices. Absent a single
smoking gun experiment the final proof of the Majorana existence 
will most likely involve a body of
additional experimental and theoretical work aimed at verifying
various aspects of the Majorana zero modes. These include the already
mentioned gap closing at the phase transition to and from the
topological phase, better theoretical understanding of the `soft gap' phenomenon
observed in the tunnelling spectroscopy, as well as several
theoretically proposed tests, argued to be less ambiguous, but so far
unrealized. These include probes of non-locality inherent to the pair of spatially separated Majorana zero modes \cite{fu1,stanescu1,tewari1,law1} and more ambitiuously direct tests of their non-Abelian exchange statistics \cite{alicea77}. 

Finally, it is likely that  in the near future 
similar experiments will be conducted in topological insulator nanowires. If
the signatures of zero modes are observed in these as well then
this would add to the evidence in favor of Majorana
fermions. Due to the less
restrictive conditions that should prevail in these systems it might
be possible to acess different regimes and observe,  for instance, the
quantized zero-bias peak conductance more easily.

Given our theoretical understanding of the problem and the mounting experimental evidence it is becoming increasingly difficult to envision a scenario in which the Majorana zero modes would not underlie at least some of the recent reports. Neverheless, 75 years after Majorana's historic prediction, the race for the unambiguous detection of these elusive particles continues.

\end{document}